\documentclass[final,5p]{elsarticle}

\usepackage{amssymb}
\usepackage{amsthm}

\usepackage{amsmath}
\usepackage{mathtools}
\usepackage{mathdots}
\usepackage{dsfont}

\usepackage{tikz}
\usetikzlibrary{external}
\usepackage{pgfplots}
\usepgfplotslibrary{groupplots}
\usepgfplotslibrary{colorbrewer}
\usepgfplotslibrary{fillbetween}
\usetikzlibrary{colorbrewer}

\usepackage{tabularx}

\graphicspath{{fig/}}

\usepackage{xcolor}

\usepackage[hidelinks]{hyperref}

\newcommand{\figref}[1]{Figure~\ref{#1}}
\newcommand{\tabref}[1]{Table~\ref{#1}}
\newcommand{\secref}[1]{Section~\ref{#1}}
\LetLtxMacro{\originaleqref}{\eqref}
\renewcommand{\eqref}{Eq.~\originaleqref}

\usepackage{booktabs} %

\newcommand{\eps}{\epsilon}
\newcommand{\state}{s}
\newcommand{\states}{\mathcal{S}}
\newcommand{\action}{a}
\newcommand{\actions}{\mathcal{A}}
\newcommand{\transition}{\mathcal{T}}
\newcommand{\sgeneraltransition}{ \transition(\state \sgiven \state_t ,\action_t )} %
\newcommand{\rewardfunc}{\mathcal{R}}
\newcommand{\reward}{r}
\newcommand{\return}{R}
\newcommand{\rewardscale}{\alpha}
\newcommand{\discount}{\gamma}

\newcommand{\traj}{\tau}
\newcommand{\loss}{\mathcal{L}}
\newcommand{\policy}{\pi}
\newcommand{\sgeneralpolicy}{ \policy_{\modelparam}(\action \sgiven  \state_t )} %
\newcommand{\probratio}{\varrho}
\newcommand{\policygradient}{g}
\newcommand{\modelparam}{\theta}
\newcommand{\valueparam}{\Theta}
\newcommand{\expectation}{\mathds{E}}
\newcommand{\ppoclipping}{\epsilon}
\newcommand{\ppoadvantage}{A}
\newcommand{\valuefunction}{V}
\newcommand{\qfunction}{Q}

\newcommand{\opt}{\star}
\newcommand{\sample}{\sim}
\newcommand{\given}{\:\big|\:}
\newcommand{\sgiven}{\,|\,}

\newcommand{\wavenumber}{k}
\newcommand{\turbenergy}{E}
\newcommand{\tfilter}[1]{\overline{#1}}

\newcommand{\eqncomma}{,}
\newcommand{\eqnperiod}{.}

\journal{}

\begin{document}

\begin{frontmatter}

\title{Deep Reinforcement Learning for Turbulence Modeling in Large Eddy Simulations}

\author[label1]{Marius Kurz\corref{cor1}}
\ead{marius.kurz@iag.uni-stuttgart.de}

\author[label2]{Philipp Offenh\"auser}
\ead{philipp.offenhaeuser@hpe.com}

\author[label1]{Andrea Beck}
\ead{beck@iag.uni-stuttgart.de}

\address[label1]{Institute of Aerodynamics and Gas Dynamics, University of Stuttgart, Pfaffenwaldring 21, 70569 Stuttgart, Germany}
\address[label2]{Hewlett Packard Enterprise (HPE), Herrenberger Straße 140, 71034  Böblingen, Germany}

\cortext[cor1]{Corresponding author}

\begin{abstract}

  Over the last years, supervised learning (SL) has established itself as the state-of-the-art for data-driven turbulence modeling.
  In the SL paradigm, models are trained based on a dataset, which is typically computed a priori from a high-fidelity solution by applying the respective filter function, which separates the resolved and the unresolved flow scales.
  For implicitly filtered large eddy simulation (LES), this approach is infeasible, since here, the employed discretization itself acts as an implicit filter function.
  As a consequence, the exact filter form is generally not known and thus, the corresponding closure terms cannot be computed even if the full solution is available.
  The reinforcement learning (RL) paradigm can be used to avoid this inconsistency by training not on a previously obtained training dataset, but instead by interacting directly with the dynamical LES environment itself.
  This allows to incorporate the potentially complex implicit LES filter into the training process by design.
  In this work, we apply a reinforcement learning framework to find an optimal eddy-viscosity for implicitly filtered large eddy simulations of forced homogeneous isotropic turbulence.
  For this, we formulate the task of turbulence modeling as an RL task with a policy network based on convolutional neural networks that adapts the eddy-viscosity in LES dynamically in space and time based on the local flow state only.
  We demonstrate that the trained models can provide long-term stable simulations and that they outperform established analytical models in terms of accuracy.
  In addition, the models generalize well to other resolutions and discretizations.
  We thus demonstrate that RL can provide a framework for consistent, accurate and stable turbulence modeling especially for implicitly filtered LES.
\end{abstract}

\begin{keyword}
  Turbulence Modeling \sep Deep Reinforcement Learning \sep Large Eddy Simulation
\end{keyword}

\end{frontmatter}

\section{Introduction}
\label{sec:introduction}

Most flows in nature and in engineering are turbulent.
Such turbulent flows are characterized by a wide range of active flow scales oftentimes spanning orders of magnitude.
Even though the governing equations of fluid motion, the Navier-Stokes equations, are known, the direct numerical simulation (DNS) of this broad range of flow scales is usually intractable.
Instead, reduced model equations are solved, which consider only the most important scales of the flow.
The most popular approaches are the Reynolds-averaged Navier-Stokes (RANS) equations, which compute the time averaged flow field, and the large eddy simulation (LES), which resolves only the most energetic flow scales in space and time.
Both approaches introduce additional terms into the governing equations on the coarse level, which embody the footprint of the non-resolved fine scales onto the resolved flow field.
Since these terms are a function of the unknown full solution, the equations remain unclosed.
Turbulence models are typically employed to approximate the unknown closure terms based on the available coarse scale data in order to close the equations.
Despite decades of research, no overall \textit{best} model has emerged yet.
Moreover, most models employ empirical model parameters that have to be tuned to the specific flow and discretization at hand.
To this end, recent advances in turbulence modeling increasingly strive to complement the established mathematical and physical modeling strategies by the emerging data-driven paradigm of machine learning (ML).

As of today, most of the research in the field of data-driven turbulence modeling is concerned with supervised learning (SL) \cite{brunton2020machine,beck2021perspective,duraisamy2019turbulence}.
In SL, artificial neural networks (ANN) are used to approximate the functional relationship between an input and an output quantity based on example data pairs.
These example data pairs are drawn from a training dataset, which in the context of turbulence modeling is typically obtained a priori from experimental or high-fidelity numerical data.
ANN comprise free parameters (\textit{weights}) which can be fitted to the dataset through optimization, which is then referred to as \textit{training} or \textit{learning}.
In principle, other functions like kernel methods \cite{wenzel2021novel} can be applied as ansatz functions as well, but ANN represent the most prominent choice.
In one of the first works in the field of ANN-based turbulence modeling, Sarghini et al. \cite{sarghini2003neural} applied an ANN as surrogate for the computation of the dynamic viscosity parameter of a mixed turbulence model in order to reduce its computational cost.
Ling et al. \cite{ling2016reynolds} proposed a novel architecture to embed Galilean invariance into their ANN-based turbulence model for RANS.
Gamahara and Hattori \cite{gamahara2017searching} applied a fully-connected ANN to predict the sub-grid stresses for LES of turbulent channel flow.
More sophisticated ANN architectures are applied for turbulence modeling in LES among others by Beck et al. \cite{beck2019deep} and Kurz and Beck \cite{kurz2022machine}, who used convolutional and recurrent ANN, respectively.

The premise of SL is that a sufficiently large dataset with defined input and output quantities is available to train the ANN on.
In the context of LES, this requires that the true closure terms have to be known in order to serve as target for the training.
This is a rather natural assumption for RANS, since here the closure terms are uniquely defined through the temporal averaging.
The same holds for LES with a predefined filter form, which is also called an explicitly filtered LES.
Since the spatial filter is known, the exact closure terms can also be computed from DNS data by applying the prescribed LES filter.
Both approaches thus provide consistent input and target quantities for training in the SL context.
However, for most applications of LES the filter form is not given explicitly.
Instead, the underresolved discretization itself acts as an implicit LES filter that separates the flow scales into resolved and non-resolved scales.
This improves the efficiency of the simulation significantly, but the form of the implicit filter induced by the discretization is typically unknown.
Moreover, the induced filter is typically non-linear and depends on the grid spacing as well as the type of discretization scheme employed.
As a consequence, the closure terms for implicitly filtered LES cannot be computed from high-fidelity DNS data, since the filter that would have to be applied is unknown.

Typically, ANN are trained on surrogate targets instead.
Such surrogate targets are oftentimes obtained by using an explicit LES filter that approximates the discretization's unknown implicit filter form.
A common example is the use of the box filter kernel as approximated LES filter for finite volume methods. %
However, this leads to severe inconsistencies between the training and actual simulations, since the targets the ANN trains on are not the correct closure terms required by the implicit LES \cite{beck2019deep}.
This can lead to inaccurate results or even unstable simulations.
Common approaches to alleviate these problems are either to project the (inconsistent) predictions onto a stable basis to ensure stability \cite{beck2019deep,maulik2019subgrid}, or to use additional regularization during training to increase the robustness of the ANN against this mismatch \cite{kurz2021investigating}.
Rasp et al. \cite{rasp2020coupled} proposed a coupled online training design, in which the pre-trained models are corrected for these inconsistencies by running the model in parallel with a high-fidelity simulation and guiding the model towards this accurate solution.
However, the SL approach for turbulence modeling in implicitly filtered LES is, from our perspective, ill-posed, since the training targets are generally unknown.

An alternative approach that alleviates these problems is the reinforcement learning (RL) paradigm.
In RL, the ML model is not trained by means of an offline training set, but learns by interacting directly with the dynamical environment itself in order to achieve some high-level target.
This allows to optimize data-driven turbulence models based on how they act in actual simulations in the context of the genuine implicit LES filter instead of training ANN on static snapshots of the flow with uncertain labels.
RL is thus conceptually opposite to SL.
In SL, the training data needs to be well-defined a priori, since each input data point requires a known \textit{true} output.
In RL, the training data is generated from the evolving system (i.e. the discretized equations) itself and the \textit{best} outputs are found by the RL algorithm in order to fulfill its overall goal. %
In this sense, RL is a much more natural way of learning in and for dynamical systems that contain a degree of uncertainty as is the case for implicitly filtered LES.

Originally, RL was especially employed for flow control tasks in numerical and experimental setups, e.g. \cite{rabault2019accelerating,rabault2019artificial,tang2020robust,fan2020reinforcement}.
More recently, Novati et al. \cite{novati2021automating} employed an actor-critic RL algorithm to derive data-driven turbulence models for homogeneous isotropic turbulence.
Moreover, Kim et al. \cite{kim2022deep} derived an RL-based model for the Reynolds stresses in turbulent channel flow and Bae and Koumoutsakos \cite{bae2022scientific} applied RL for wall-modeling in turbulent channel flow.

Based on these encouraging results, we contribute in this paper the following.
We apply the novel RL framework Relexi\footnote{\url{https://github.com/flexi-framework/relexi}} proposed in \cite{kurz2022deep,kurz2022relexi} to derive data-driven turbulence models for implicitly filtered LES with a modern high-order discontinuous Galerkin (DG) scheme.
While this discretization choice serves to demonstrate the suitability of our method for modern, high-order discretizations of practical relevance, it also introduces a significant difficulty in the modeling process.
The induced a priori filter form of DG is an element-local $L_2$-projection onto the polynomial basis of the DG method.
However, since the elements are coupled via the numerical fluxes across the element faces, the resulting hybrid operator of the DG method introduces a complex filter form that typically has a non-linear kernel.
We use forced homogeneous isotropic turbulence as a representative canonical test case for training, and learn an optimal, time- and space-dependent eddy viscosity.
We employ the proximal policy optimization (PPO) algorithm for training, which achieves state-of-the-art results in many RL tasks, while being relatively straight-forward to implement.
The architecture of the policy network is based on three-dimensional convolutional layers, which allows to incorporate the flow state in the vicinity of the investigated point into the prediction.
However, the inputs and outputs of the policy are strictly element-local, which means that no information of the global flow state is required as input by the policy.
The locality of the input is a crucial property for practical applications, since global flow statistics are generally expensive to obtain during simulations and are also oftentimes ill-defined therein, especially on more complex computational domains.
This improves on the policy inputs employed by Novati et al. \cite{novati2021automating}, who used additionally the global dissipation rate and the global energy spectrum as inputs for the policy.
We demonstrate that the derived models are long-term stable and that they outperform established analytical LES models in terms of accuracy.
In a last step, we show that the trained models generalize well to different resolutions and to flows at higher Reynolds numbers.

The paper is organized as follows.
\secref{sec:rl} gives a brief outline of the RL paradigm and the PPO algorithm.
In \secref{sec:turbulence}, the task of turbulence modeling is introduced and formulated as an RL problem, which can be solved with the proposed RL framework.
The results in \secref{sec:results} give details on the training results and assess the performance and the properties of the trained RL-based models.
\secref{sec:conclusion} concludes the paper.

\section{Reinforcement Learning}
\label{sec:rl}

\subsection{A Brief Introduction}
\label{sec:rl_intro}

\begin{figure}
  \centering
  \includegraphics[width=0.7\linewidth]{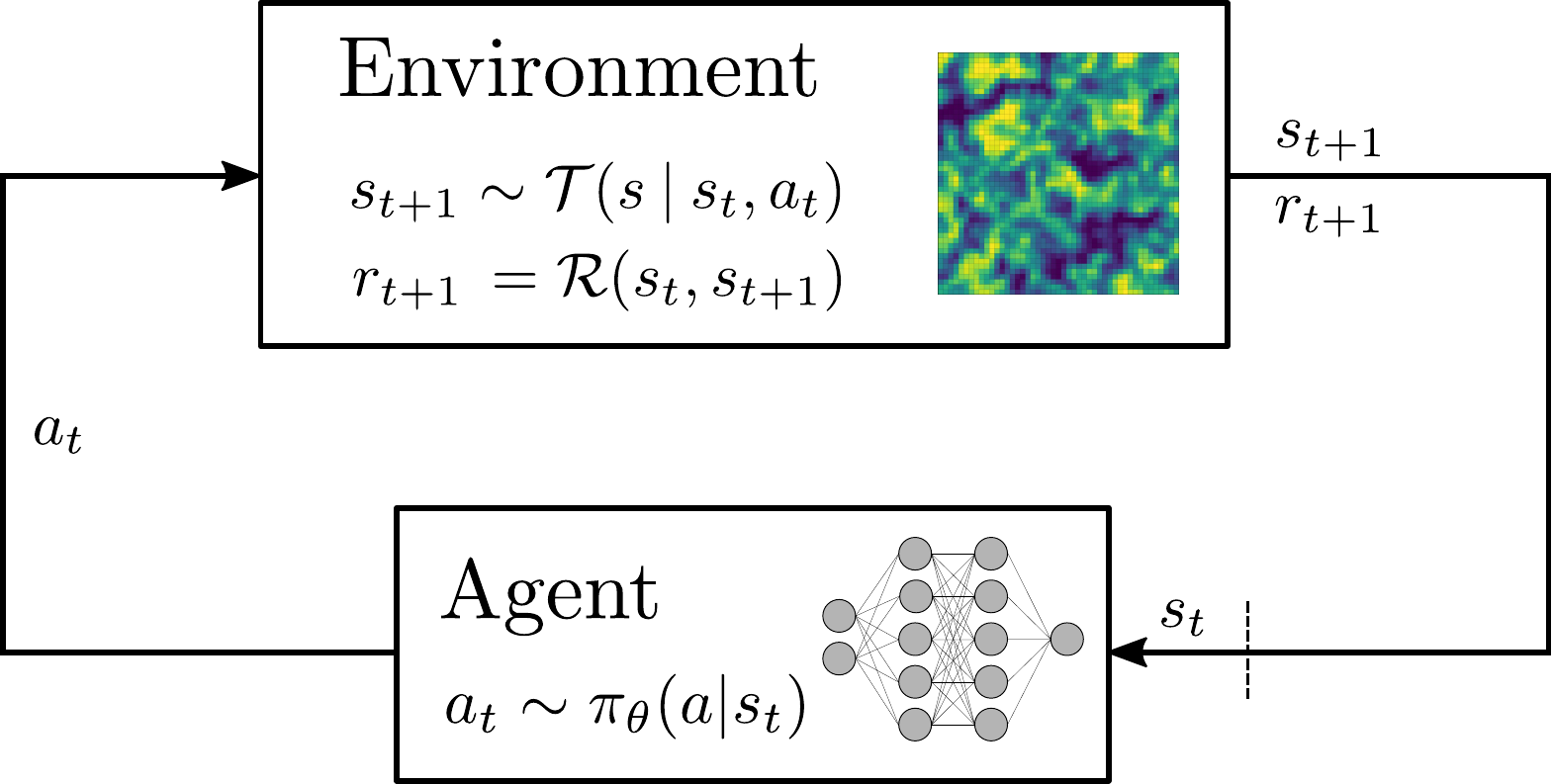}
  \caption{Schematic of the Markov decision process. The agent performs an action $a_t$ based on its policy $\sgeneralpolicy$ to interact with the environment. Consequently, the environment transitions into a new state $\state_{t+1}$ according to its transition function $\sgeneraltransition$. The resulting reward $\reward_{t+1}$ is specified by the reward function \mbox{$r_{t+1}=\rewardfunc\left(s_t,\state_{t+1}\right)$}, which is used to quantify how desirable a given state transition is. Starting from an initial environment state $\state_0$ this process is repeated until a final state $\state_n$ is reached. }
  \label{fig:MDP}
\end{figure}

In contrast to the supervised learning approach, reinforcement learning trains an agent by letting it interact with a dynamical environment in order to achieve a pre-defined goal.
This has the advantage that the dynamics of the environment are incorporated into the training process directly by design.
The interplay of the agent and the environment can be framed as a Markov decision process (MDP), as illustrated in \figref{fig:MDP}.
In a MDP, the environment is characterized by its current state $s_t\in\states$.
The agent observes that state and can perform a suitable action $a_t\in\actions$.
Here, $\states$ is the set of all possible environment states and $\actions$ is the set of all possible actions that can be performed by the agent.
The agent's actions are determined by its policy $\sgeneralpolicy$ that states which action the agent should perform given the environment's current state $\state_t$.%
\footnote{
  To keep the notation short, we use the abbreviated notation of $\sgeneralpolicy$ for $\policy_{\modelparam}(\action\sgiven\state=\state_{t})$ which describes the conditional probability of choosing action $\action$ given the current state $\state_t$.
The same holds for the transition function $\transition\left(\state'\sgiven\state=\state_{t},\action=\action_{t}\right)$, which will be abbreviated as $\sgeneraltransition$.
}.
The policy can be of any functional form, but for deep RL, the policy is typically an ANN with parameters $\modelparam$.
The agent's action causes the environment to change is state.
This new state is prescribed by the environment's transition function $\state_{t+1}\sample\sgeneraltransition$, which thus encodes the environment's dynamics.
With the state transition, the agent receives a reward $\reward_{t+1}$ that is determined by the reward function $\reward_{t+1}=\rewardfunc(\state_{t},\state_{t+1})$ and quantifies how desirable a certain state transition is.
The new state $s_{t+1}$ is then again observed by the agent, which performs another action $a_{t+1}$ as prescribed by its current policy.
Starting from some initial state $\state_0$ and performing actions until a final state $\state_n$, this results in a trajectory of states, actions and rewards termed an episode:
\begin{equation}
  \traj = \left\{ \left(\state_0,\action_0,\reward_1\right),\left(\state_1,\action_1,\reward_2\right),\:......\;,\left(\state_{n-1},\action_{n-1},\reward_{n}\right),\state_n\right\} \eqnperiod
  \label{eq:trajectory4}
\end{equation}
The goal of an RL algorithm is to establish an optimization problem that allows to find the optimal policy $\policy^{\opt}$, which maximizes the expected return along an episode
\begin{equation}
  \return(\tau) = \sum_{t=1}^{n} \discount^t r_{t} \eqnperiod
\end{equation}
Here, $\discount$ denotes the discount factor that can be used to balance the importance of short-term and long-term rewards.
In deep RL, finding the optimal policy is equivalent to finding the set of optimal model parameters $\modelparam^{\opt}$ for the employed ANN.

For each state $\state$, the \textit{state-value function} describes the total return which can be expected starting from state $\state_t$ and following a specific policy $\policy$ from there on.
This state-value function can thus be written as
\begin{equation}
  \valuefunction^\policy \left(\state\right) = \expectation\left[\sum_{k=0}^\infty \discount^k r_{t+k}\given\state_t=\state\right] \eqnperiod
  \label{eq:valuefunction}
\end{equation}
Similarly, an \textit{action-value function} or \textit{Q-function} can be determined, which gives the expected return when starting from state $\state_t$ performing action $\action_t$ and following the policy $\policy$ from there on, which reads
\begin{equation}
  \qfunction^\policy \left(\state,\action\right) = \expectation\left[\sum_{k=0}^\infty \discount^k r_{t+k}\given\state_t=\state, \action_t=\action\right] \eqnperiod
  \label{eq:qfunction}
\end{equation}
Based on \eqref{eq:valuefunction} and \eqref{eq:qfunction}, one can define the \textit{advantage function}
\begin{equation}
  \ppoadvantage^\policy \left(\state,\action\right) =  \qfunction^\policy \left(\state,\action\right) - \valuefunction^\policy \left(\state\right) \eqncomma
  \label{eq:advantagefunction}
\end{equation}
which quantifies whether taking action $a_t$ in state $s_t$ increases or decreases the expected return in comparison to performing the action prescribed by the current policy.

Solving a given problem with RL thus requires that the problem is casted into an MDP by a domain expert, i.e. defining the environment's possible states $\states$, its transition function $\sgeneraltransition$, the agent's action space $\actions$, the reward function $\rewardfunc(s_t,s_{t+1})$ and, finally, the ANN architecture used for the policy $\sgeneralpolicy$.
With these definitions in place, a suitable RL algorithm can be applied to find a favorable policy.
Each distinct RL algorithm prescribes how interactions of the agent and the environment are collected and how this sampled experience can be used to optimize the policy such that the expected future return is increased.
RL algorithms differ for instance in terms of sample efficiency and whether they allow for continuous state and action spaces.
In the following, we use proximal policy optimization (PPO) as our RL algorithm of choice, which belongs to the class of policy gradient methods.

\subsection{Policy Gradient Methods}
\label{sec:vpg}

The key idea of policy gradient methods is to optimize the policy directly instead of learning the Q-function in \eqref{eq:qfunction} and inferring the policy implicitly from it.
To this end, policy gradient methods derive a gradient estimator that gives the direction in which the model parameters $\modelparam$ have to be changed in order to increase the expected return.
Given this gradient, the model parameters can be updated with a suitable gradient-ascent algorithm.
Following the \textit{policy gradient theorem}, see e.g. \cite{sutton2018reinforcement}, the gradient estimator can be written as
\begin{equation}
  \policygradient = \expectation\Big[\qfunction^{\policy}\left(\state,\action\right) \,\nabla_{\modelparam}\log \policy_{\modelparam}\left(\action\sgiven\state \right) \Big],
  \label{eq:pg_gradient}
\end{equation}
and can be obtained by differentiating the corresponding loss function
\begin{equation}
  \loss^{VPG}(\modelparam) = \expectation\Big[ \qfunction^{\policy}\left(\state,\action\right) \,  \log \policy_{\modelparam}\left(\action\sgiven\state \right)\Big],
  \label{eq:pg_loss}
\end{equation}
with respect to $\modelparam$.
Since the policy and the environment are in general stochastic, the gradient estimator for the optimization is defined by means of the expectation $\expectation\left[\cdot\right]$.
However, obtaining the exact expectation is prohibitive for practical applications.
Instead, the gradient estimator is approximated by sampling $N$ trajectories of experience on the current policy and computing the approximated gradient as mean over the sampled trajectories
\begin{equation}
  \hat{\policygradient} = \frac{1}{N}\sum_{i=1}^N\left[\return\left(\traj^{(i)}\right) \ \sum_{t=0}^n  \nabla_{\modelparam}\log \policy_{\modelparam}\left(\action_t\sgiven\state_t \right)\right] \eqnperiod
  \label{eq:pg_approx_gradient}
\end{equation}
Here, the discounted return along a trajectory $\return(\traj)$ is used as an approximation to the exact Q-function.
Interestingly, the dynamics of the environment, i.e. its transition function $\sgeneraltransition$, do not appear in the gradient estimator or in the overall optimization formulation.
Instead, the dynamics of the environment are incorporated implicitly by the sampled experience.
This avoids to differentiate the environment dynamics with respect to the policy parameters, which is infeasible for most tasks.

The training process of a policy gradient method then works as follows.
First, multiple episodes of experience are sampled with the current policy.
Based on this experience, the policy can be optimized in a second step with the gradient estimator and a suitable gradient-ascent algorithm.
These two steps are then repeated, until the policy has converged.
These building blocks form the vanilla policy gradient (VPG) method.

\subsection{Proximal Policy Optimization}
\label{sec:ppo}

The proximal policy optimization (PPO) method \cite{schulman2017proximal} introduces several improvements over the original vanilla policy gradient (VPG) method to improve the stability of the training.
For a clear and concise summary of the PPO algorithm, we recommend \cite{notter2021hierarchical}.

The first major improvement of PPO is to reduce the variance of the gradient estimator.
This reduces the amount of samples required for an accurate approximation of the gradient or a better gradient estimator for a given amount of samples.
To this end, a baseline can be added to the gradient estimator in \eqref{eq:pg_approx_gradient}, which has been shown to not introduce a bias.
A natural choice is to replace the Q-function by the advantage function from \eqref{eq:advantagefunction}.
However, the advantage function relies on the state-value function $\valuefunction^{\policy}(s)$, which is typically unknown.
Therefore, the PPO algorithm uses an additional ANN $\hat{\valuefunction}_{\valueparam}(s)$ with weights $\valueparam$, which is trained to approximate the state-value function.
Moreover, the return along the trajectory, which approximates the Q-function, is replaced by a \textit{return-to-go} $R_t(\tau) = \sum_{k=t}^n \gamma^{k-t} r_k$ such that each action is only associated with reward that is collected after the action is taken.
The approximate advantage function at step $t$ thus reads
\begin{equation}
  \hat{\ppoadvantage}_t = \left(\sum_{k=t}^n \gamma^{k-t} r_k\right) - \hat{\valuefunction}_{\valueparam}(s_t).
  \label{eq:ppo_advantage}
\end{equation}

A major drawback of VPG is that the training with the VPG method is often found to be unstable, since the policy updates can become arbitrarily large.
Large policy updates imply the risk of deteriorating the policy's performance in a single update step if the gradient estimator is not sufficiently accurate or the step size is too large.
The PPO method increases the stability of the training process by constraining the maximum change of the policy in a single step. %
Schulman et al. \cite{schulman2017proximal} propose two different approaches to limit the updates of the policy.
Firstly, a penalty term can be added to \eqref{eq:pg_loss} based on the Kullback-Leibler divergence between the old and the new policy.
This introduces the incentive to avoid large changes in the policy in a single training step.
The other approach is to replace the loss function in \eqref{eq:pg_loss} by a clipped surrogate objective
\begin{equation}
  \loss^{CLIP}(\modelparam) = \expectation\left[\min\left( \probratio\left(\modelparam\right)\hat{\ppoadvantage},\mathrm{clip}\left(\probratio\left(\modelparam\right),1-\eps,1+\eps\right)\hat{\ppoadvantage}   \right)   \right],
  \label{eq:clipping_loss}
\end{equation}
with $\eps$ as a hyperparameter and with $\probratio(\modelparam)$ as the probability ratio
\begin{equation}
  \probratio\left(\modelparam\right)=\frac{\policy_{\modelparam}\left(\action\sgiven\state\right)}{\policy_{\modelparam_{old}}\left(\action\sgiven\state\right)},
  \label{eq:prob_ratio}
\end{equation}
which describes the change between the old and the new policy.
In \eqref{eq:clipping_loss}, the clip function clips the probability ratio $\probratio(\modelparam)$ to the interval $\left[1-\epsilon,1+\epsilon\right]$ to limit the change of the policy in a single training step.
Similarly to \eqref{eq:pg_approx_gradient}, the expectation in \eqref{eq:clipping_loss} is then approximated by sampling trajectories with the current policy.

\section{Data-Driven Turbulence Modeling}
\label{sec:turbulence}

\subsection{Governing Equations}
\label{sec:equations}
The temporal evolution of compressible, viscous flows is described by the Navier-Stokes equations, which can be written as
\begin{equation}
  U_t + \nabla_x\cdot \left(F^c - F^v\right) = 0,
  \label{eq:navier_stokes}
\end{equation}
with $U=\left(\rho, \rho v_1, \rho v_2, \rho v_3, \rho e \right)^T$ as the vector of conserved quantities comprising density, the three-dimensional momentum vector and energy, respectively.
Here, $(\cdot)_t$ indicates the differentiation with respect to time and the nabla operator $\nabla_x$ the differentiation with respect to the three Cartesian coordinates $x_i$ with $i=1,2,3$.
The convective flux $F^c$ and the viscous flux $F^v$ with columns $i=1,2,3$ can be written as 
\begin{equation}
  F_i^c =
  \left(
  \begin{array}{c}
    \rho v_i \\
    \rho v_1 v_i + \delta_{1i} p\\
    \rho v_2 v_i + \delta_{2i} p\\
    \rho v_3 v_i + \delta_{3i} p\\
    \rho e v_i + p v_i
  \end{array}
  \right)
  ,\;
  F_i^v =
  \left(
  \begin{array}{c}
    0 \\
    \sigma_{1i}\\
    \sigma_{2i}\\
    \sigma_{3i}\\
    \sigma_{ij}v_j-q_i
  \end{array}
  \right) ,
  \label{eq:fluxes_written_out}
\end{equation}
where $p$ denotes the pressure and $\delta_{ij}$ is the Kronecker delta.
The three-dimensional velocity vector is given by $v =(v_1,v_2,v_3)^T$.
The stress tensor $\sigma_{ij}$ and the heat flux $q_i$ can be written as
\begin{align}
  \sigma_{ij} &= \mu\left(\frac{\partial v_i}{\partial x_j}+\frac{\partial v_j}{\partial x_i}-\frac{2}{3}\delta_{ij}\frac{\partial v_k}{\partial x_k}\right) ,
  \label{eq:stress_tensor}\\
  q_i &= - \kappa\:\frac{\partial T}{\partial x_i},
  \label{eq:heat_flux}
\end{align}
with $\mu$ as the viscosity and $\kappa$ as the heat conductivity.
The equations are closed with the ideal gas assumption, which yields the equation of state as
\begin{equation}
  p = \left(\frac{c_p}{c_v}-1\right) \left(\rho e-\frac{\rho}{2}\left(v_1^2+v_2^2+v_3^2\right)\right),
\end{equation}
with $c_p$ and $c_v$ as the specific heats.
In the following, the Navier-Stokes equations are always solved in the incompressible limit with a Mach number of $\mathrm{Ma}=0.1$.

\subsection{Forced Homogeneous Isotropic Turbulence}

\begin{figure}
  \begin{minipage}[t]{0.443\linewidth}
    \includegraphics[width=\linewidth]{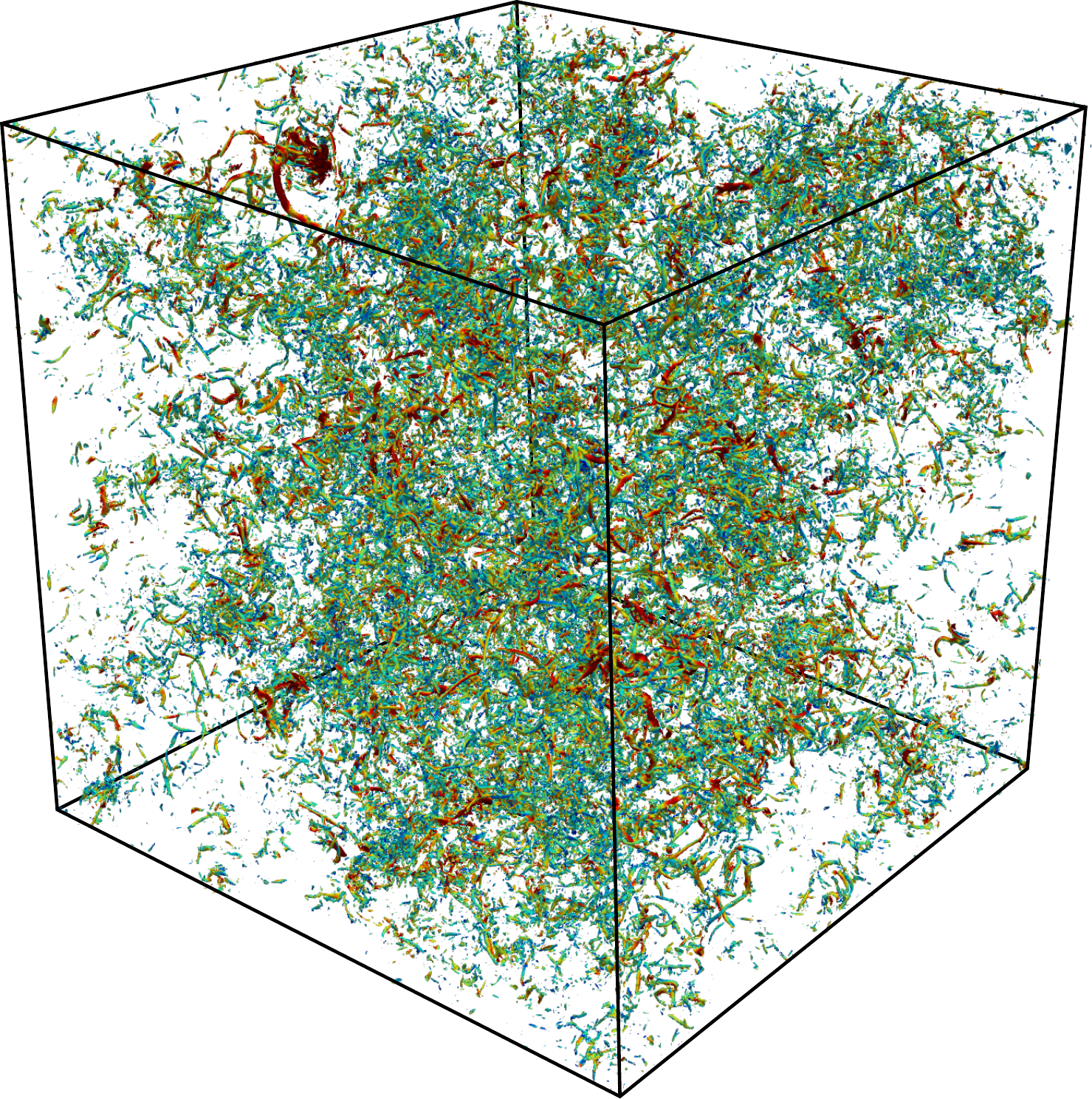}%
  \end{minipage}
  \hfill
  \begin{minipage}[t]{0.53\linewidth}
    \includegraphics[width=\linewidth]{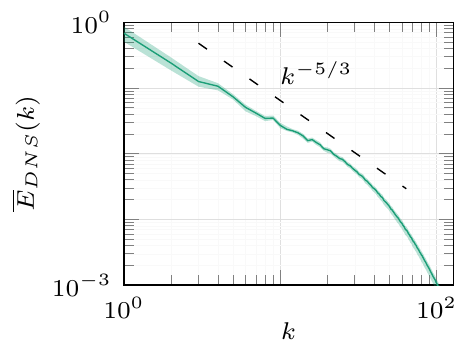}%
  \end{minipage}
  \caption{Instantaneous flow field of the HIT test case visualized with iso-surfaces of the Q-criterion colored by the velocity magnitude (left) and the mean spectrum of the kinetic energy for the DNS of a forced HIT over the wavenumbers $k$ (right). The shaded area indicates the maximum and minimum observed energy of the corresponding wavenumber.}
  \label{fig:HIT}
\end{figure}

Homogeneous isotropic turbulence (HIT) is a canonical test case for turbulent flows and can be considered as \textit{turbulence-in-a-box}.
The HIT test case describes freely evolving turbulence without an imposed mean flow, influences of walls or any external driving forces.
The computational domain is typically a cube with periodic boundary conditions, as illustrated in \figref{fig:HIT}, which is discretized by an equidistant Cartesian mesh.
The domain is initialized with an initial velocity field that obeys a given turbulence spectrum and is divergence-free.
In this work, the initial flow state was obtained by Rogallo's procedure \cite{Rogallo1981}.
Over time, the flow then produces a turbulent energy spectrum as shown in \figref{fig:HIT}.
In turbulence, energy is transported in the mean from the large scales to the small scales, where the energy is dissipated by the viscosity of the fluid.
In the absence of large scale shear and due to this dissipation mechanism, the velocity fluctuations decrease over time and tend towards zero.
Therefore, this flavor of the HIT test case is also referred to as decaying HIT.
However, this transient behavior causes the flow and the turbulent statistics to change continuously over time.

Different forcing strategies are proposed in literature to inject the dissipated energy back into the system and thus sustain the turbulent flow.
This allows to obtain a turbulent flow with stationary statistics.
For methods with a globally defined solution basis, the forcing can be added in the low modes only, while the higher wavenumbers remain unaffected.
For discretizations with a local basis with compact support, e.g. finite-volume or discontinuous Galerkin schemes, the global Fourier modes of the solution are generally unknown and costly to evaluate.
Instead, a nodal forcing is applied, which acts directly on the solution points instead of global wavenumbers.
Since in this approach the forcing is not limited to the low wavenumbers, the forcing adds energy across the whole spectrum.
This causes the slope in \figref{fig:HIT} to slightly deviate from its theoretical $k^{-5/3}$ trend, since this influx of energy into the high wavenumber modes is not optimal. %
We point out however that, first, such a nodal forcing is often used in element-based discretization schemes and, second, the results in \figref{fig:HIT} indicate that the unwanted effects are minimal.

In this work, we apply the linear isotropic forcing method proposed by Lundgren in \cite{lundgren2003linearly} and analyzed further in \cite{de2015anisotropic}.
Here, a forcing term $f$ is added to \eqref{eq:navier_stokes}, which yields
\begin{equation}
  U_t + \nabla_x\cdot \left(F^c - F^v\right) = f.
  \label{eq:forced_navier_stokes}
\end{equation}
For isotropic forcing, the forcing is assumed to be parallel to the current momentum vector, which gives
\begin{equation}
  f = Q \: 
  \left(
  \begin{array}{c}
    0 \\
    \rho v\\
    0
  \end{array}
  \right),
\end{equation}
where $Q$ is a scalar that quantifies the difference between the current kinetic energy in the flow $E=\frac{\rho}{2} (v\cdot v)$ integrated over the domain and the prescribed target value.
In our implementation, a single forcing parameter $Q$ is employed for the whole flow domain.

\subsection{Large Eddy Simulation}
\label{sec:les}

\begin{figure}[tb]
  \centering
  \includegraphics[width=0.99\linewidth]{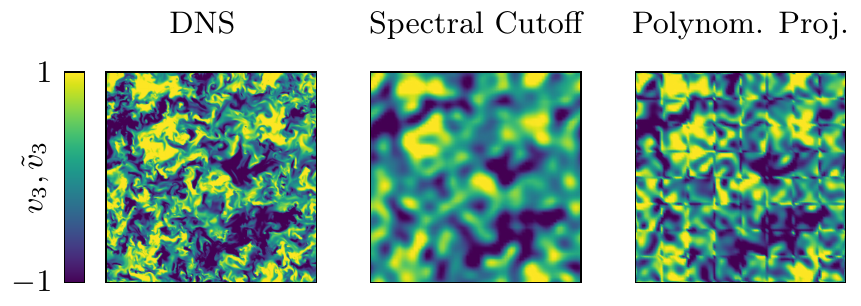}
  \caption{Slices of the instantaneous velocity field of homogeneous isotropic turbulence. The LES flow fields are obtained from the DNS field (left) by applying a spectral cutoff filter (middle) or a projection filter onto a piecewise polynomial basis (right). The latter explicit LES filter can be seen as a rough approximation of the unknown implicit filter form of the DG scheme. More details on the LES filters are given in \cite{kurz2022machine}.}
  \label{fig:les_filter}
\end{figure}

For most flows, it is computationally intractable to resolve all length scales present in the flow.
Instead, the simulation resolves only the largest flow scales, which contain the majority of the kinetic energy.
This approach is referred to as large eddy simulation (LES).
This corresponds to applying a low-pass filter $\tilde{(\cdot)}$ to the governing equations in \eqref{eq:navier_stokes} and solving the resulting evolution equation for the coarse-scale solution $\tilde{U}$.
In \figref{fig:les_filter}, we apply known filter kernels in an a priori manner to visualize the resulting solution fields.
Due to the non-linearity of the governing equations, the filtering operation introduces additional terms into the equation, which describe the effect of the non-resolved fine scales onto the resolved large scales.
These closure terms rely on the full solution $U$ and are thus generally unknown causing the closure problem of LES.
Therefore, a suitable turbulence model is employed that attempts to approximate the unknown closure terms based on the coarse-scale solution $\tilde{U}$.

From \figref{fig:les_filter} we can already appreciate the fact that the associated closure terms will be a function of the chosen filter.
In practice however, the filtering operation is typically not given by means of an explicit filter function.
Instead, the underresolved discretization acts as an implicit LES filter.
While this allows to incorporate all wavenumbers within the resolution limits of the underlying discretization into the simulation and thus improves the efficiency of the LES, the form of the LES filter $\tilde{(\cdot)}$ is generally unknown.
As a consequence, the closure terms for implicitly filtered LES cannot be computed, even if the full solution $U$ is available, since this would require to apply the unknown implicit LES filter, i.e. an application of the spatial and temporal operators.
This makes model development difficult, since the models cannot be optimized by simply fitting them to the \textit{exact} closure terms computed from high-fidelity data.

A common modeling strategy is to mimic the energy transport from the large to the small scales in turbulent flows by introducing a turbulent viscosity.
For instance, Smagorinsky's model \cite{smagorinsky1963general} computes this viscosity as 
\begin{equation}
  \mu_t = \rho\left(C_s\,\Delta\right)^2 \sqrt{2\,\tilde{S}_{ij}\,\tilde{S}_{ij}} \quad \text{with}\quad \tilde{S}_{ij}=\frac{1}{2}\left(\frac{\partial \tilde{v}_i}{\partial x_j}+\frac{\partial \tilde{v}_j}{\partial x_i}\right),
  \label{eq:smago}
\end{equation}
where $\tilde{S}_{ij}$ is the resolved rate-of-strain tensor, $\Delta$ is the filter width of the LES filter and $C_s$ is a model parameter, which has to be tuned manually to specific flows and discretizations.
The eddy-viscosity methodology has the advantage that the LES equations for the coarse-scale solution 
\begin{equation}
  \tilde{U}_t + \nabla_x\cdot \left(\tilde{F}^c - \tilde{F}^v_{turb}\right) = \tilde{f},
  \label{eq:les_equation}
\end{equation}
look identical to forced Navier-Stokes equations in \eqref{eq:forced_navier_stokes}.
Only the viscous flux $\tilde{F}^v_{turb}$ is modified slightly by adding the turbulent viscosity $\mu_t$ to the physical one.
Another common approach is the implicit modeling strategy.
Here, it is acknowledged that the employed discretization adds numerical dissipation to the system which can be interpreted as an implicit turbulence model.

In the standard Smagorinsky model (SSM), the parameter $C_s$ is constant in the computational domain and has to be chosen a priori.
So-called dynamic models alleviate these restrictions by adapting their model parameters based on the current flow state dynamically in space and time.
This concept gives rise to the dynamic Smagorinsky model (DSM), which is detailed in \ref{app:dynsmago}.
In the following, we strive to enhance the Smagorinsky model by means of our RL framework by training an agent to dynamically adapt the model coefficient in space in time during the LES, i.e $C_s=C_s(x_i,t)$.

\subsection{Reinforcement Learning for Turbulence Modeling}

\begin{table*}[htb!]
  \begin{tabularx}{1.0\linewidth}{p{0.125\textwidth}p{0.175\textwidth}X}
    \toprule
    Symbol & Meaning & Turbulence Modeling \\
    \midrule
    $\states$                  & Environment states    & Current flow state of the LES $\tilde{U}$ from which the policy's inputs can be computed.\\
    $\actions$                 & Agent's action space  & Elementwise parameter for Smagorinsky's model within $C_s \in[0,0.5]$ (\figref{fig:policy}).\\
    $\sgeneralpolicy$          & Agent's policy        & CNN-based architecture with elementwise inputs and outputs (\figref{fig:policy}).\\
    $\sgeneraltransition$      & Transition function   & Integration of LES equations, \eqref{eq:les_equation}, for current predictions of $C_s$.\\
    $\rewardfunc(s_t,s_{t+1})$ & Reward function       & Based on error in turbulent energy spectra compared to DNS solution, \eqref{eq:reward}.\\
    \bottomrule
  \end{tabularx}
  \caption{Definition of the major building blocks for the formulation of turbulence modeling as a Markov decision process, as given in \figref{fig:MDP}.}
  \label{tab:rl_turb}
\end{table*}

\begin{figure*}[htb]
  \centering
  \includegraphics[width=0.99\textwidth]{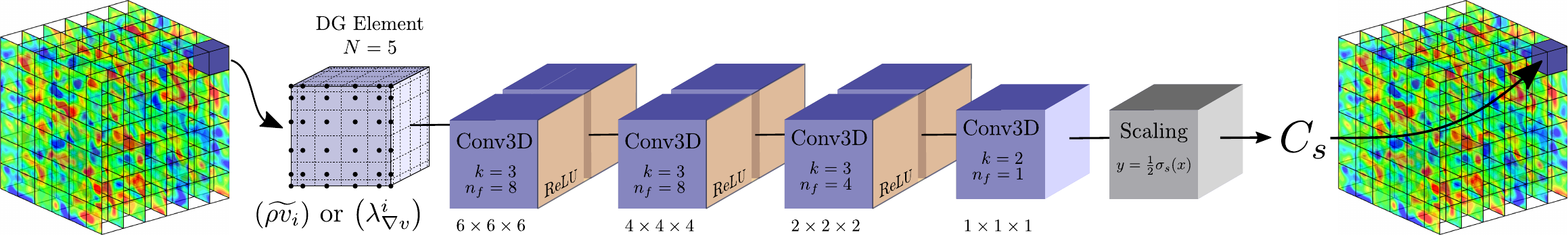}
  \caption{Network architecture of the CNN-based policy network for $N=5$. The inputs of the network are either the momentum field $(\widetilde{\rho v}_1,\widetilde{\rho v}_2,\widetilde{\rho v}_3)^T$ or the five invariants of the velocity gradient tensor $\lambda_{\nabla v}^i$ in a single DG element with given $N$, for which the distribution of interpolation points is shown exemplarily. The network comprises several three-dimensional convolutional layers (Conv3D) with the corresponding kernel sizes $k$ and the number of filters per layer $n_f$. The output sizes in the dimensions of convolution are given below each layer. The first layer retains the input dimension by means of zero-padding, while for the other layers no padding is employed in order to retain a scalar output $C_s$ per element. The scaling layer applies the sigmoid activation function $\sigma_s(x)$ in order to scale the network output to $C_s\in [0,0.5]$ and can be interpreted as the activation function of the last layer.}
  \label{fig:policy}
\end{figure*}

The problem of turbulence modeling has to be framed as an MDP in order to solve it within the RL framework, as already discussed in \secref{sec:rl}.
In the following, we use LES of forced HIT as the training environment for the agent.
The agent interacts with the environment by adapting the model coefficient of \eqref{eq:smago} dynamically in space in time for each element. %
We use a policy based on convolutional ANN (CNN) to account for the non-locality of turbulence.
To this end, the policy takes the flow state in a single element as input and predicts a single $C_s$ parameter for the respective element as output, as illustrated in \figref{fig:policy} for $N=5$.
The sigmoid function is applied to the output of the ANN to scale the prediction to $C_s\in[0,0.5]$.
Since only $C_s^2$ is used in Smagorinsky's model in \eqref{eq:smago}, we avoid negative predictions to preserve a monotonous relationship between the prediction and the actual eddy-viscosity introduced in the flow.
This choice was made to stay true to the original idea of Smagorinsky's model.
The upper limit is selected generously, while retaining the predictions in a sensible range to accelerate training.%
\footnote{It was verified for selected training configurations that the agent reaches similar levels of accuracy if the limits are omitted, i.e. $C_s\in(-\infty,\infty)$. However, the agent requires much more training iterations in this case than with the scaling applied.}

The states $\states$ are represented by the vector of conserved quantities $\tilde{U}$ on the coarse LES mesh, which gives a complete description of the current flow state.
Based on this state, two different sets of input quantities are investigated for the policy.
First, we use the elementwise three-dimensional momentum field as input, i.e. the vector $(\widetilde{\rho v}_1,\widetilde{\rho v}_2,\widetilde{\rho v}_3)^T$.
While this is a common choice in ANN-based turbulence modeling, the momentum field lacks invariance with respect to scaling and rotation.
To this end, we also investigate the input features proposed by Novati et al. \cite{novati2021automating}, who used as input quantities the five invariants of the velocity gradient tensor $\lambda_{\nabla v}^i$, as given by Pope in \cite{pope1975more}, to embed Galilean invariance into their policy.

It is important to stress that in contrast to previous works, our policy acts on \textit{local} inputs only.
This means that the predictions for each element are based solely on the flow state inside of this respective element.
The predictions for each individual element are thus independent from the remaining flow field and do not require any information about the global flow state.
This makes this approach more suitable for realistic applications, where global flow statistics are generally very expensive to retrieve during runtime.
Moreover, models which rely on global flow statistics as inputs cannot be evaluated on flows where these statistics cannot be obtained in a straight-forward manner, e.g. due to the geometric complexity of the computational domain or where a global statistic simply does not exist.
The value network is designed analogously to the policy network, but employs two additional fully-connected layers, comprising 16 and a single neuron, respectively, to combine the elementwise predictions for the whole flow state to a single prediction for the state-value function.
This RL design can be interpreted as a multi-agent RL approach, where each element employs its own agent, but all agents share their weights.
Given the predictions of the policy, i.e. the actions of the agent, the new state of the environment is obtained by integrating the LES equations in time for $\Delta t_{RL}$. 

As overall optimization target of the RL task, we define the error in the spectrum of turbulent kinetic energy between the instantaneous LES $E_{LES}$ and the target spectrum $\overline{E}_{DNS}$.
In this work, we used the time-averaged spectrum of a previously computed DNS.
However, the RL framework allows to impose any high-level statistic as training objective.
To this end, also data reported in the literature, experimental data or even physical properties like the $k^{-5/3}$ energy decay in the inertial range could be used directly as optimization targets. 
Given these energy spectra, we compute the mean squared error for the wavenumbers $k$ up to $k_{max}$.
Here, $k_{max}$ is the maximum wavenumber considered for optimization and thus depends on the employed resolution of the LES.
Finally, we use the exponential function to normalize the resulting reward to the range $[-1,1]$, since a normalized reward can improve the training speed.
The reward function thus reads
\begin{equation}
  \rewardfunc(\state) = 2\:\exp\left(-\frac{1}{\rewardscale\:\wavenumber_{max}}\sum_{\wavenumber=1}^{\wavenumber_{max}}\left(\frac{\overline{\turbenergy}_{DNS}(\wavenumber)-\turbenergy_{LES}(\wavenumber)}{\tfilter{\turbenergy}_{DNS}(\wavenumber)}\right)^2\right)-1,
  \label{eq:reward}
\end{equation}
with $\alpha$ as a scaling factor.
This scaling factor controls the difficulty of the training task by balancing between providing the maximum reward for any action of the policy for $\alpha \rightarrow \infty$ and demanding impossible accuracy in the spectrum to escape the negative limit of the exponential function for $\alpha \rightarrow 0$.
It seems important to stress that the energy spectra only have to be computed to evaluate the reward function during training and are not required if the trained policy is later applied in practical simulations.
The proposed formulation of turbulence modeling as an MDP is summarized in \tabref{tab:rl_turb}.

A major advantage of the broader learning paradigm of RL in comparison to the common SL approach is that RL allows to incorporate a much broader range of prior information into the training.
The optimization target for the training, which is defined via the reward function, can be based on DNS data, experimental data, physical first principles or properties of the underlying discretization.
For SL, any change in the posed learning task, e.g. changing the input and output quantities for the ANN, would require to recompute the training dataset from DNS data.
This poses significant challenges, since it requires to store and process large amounts of time-resolved DNS data.

\subsection{Computational setup}
\label{sec:relexi}

\begin{table}[htb]
  \centering
  \begin{tabular}{lrrrr}
    \toprule
    Name   & $N$ & \#Elems & $k_{max}$ & $\alpha$ \\
    \midrule
    24 DOF &   5 &   $4^3$ &         9 &      0.4 \\
    32 DOF &   7 &   $4^3$ &        12 &      0.4 \\
    36 DOF &   5 &   $6^3$ &        13 &      0.4 \\
    48 DOF &   5 &   $8^3$ &        16 &      0.4 \\
    \bottomrule
  \end{tabular}
  \caption{Investigated configurations for the LES environments. The names are derived from the number of degrees of freedom (DOF) in each direction used for the simulation, which can be computed as the number of elements in each direction times $(N+1)$.}
  \label{tab:les_configs}
\end{table}

In RL, the training of the agent is an iterative process which does not require the perfect target quantities to be known, but finds suitable targets by itself in order to fulfill the overall goal defined by the reward function.
This means that no DNS data is required to compute input and output data pairs for a training dataset.
Instead, a number of LES runs has to be computed repeatedly to sample interactions between the agent and the environment.
Thus, like in most ML methods, training requires considerable hardware resources for running the LES and training the agent on the collected experience.
In this work, we use the Relexi framework proposed in \cite{kurz2022deep,kurz2022relexi} to train the agent by using the parallel computing resources provided by today's high-performance computing (HPC) systems.
The Relexi framework implements the RL training loop by means of the TensorFlow library \cite{abadi2016tensorflow} and its RL extension TensorFlow-Agents \cite{TFAgents}.
The framework allows to couple external flow solvers efficiently on high-performance computing systems with the SmartSim library \cite{partee2021using}, which is used to manage the simulations runs and implements the communication between the main application and the external solver.
The LES environments are simulated with the HPC flow solver FLEXI \cite{krais2021flexi}.
In each training iteration, Relexi starts several FLEXI simulations to sample interactions of the current policy with the LES environments.

With FLEXI we employ a high-order discontinuous Galerkin (DG) discretization of the compressible Navier-Stokes equations using a kinetic energy preserving split formulation of the fluxes on Legendre-Gauss-Lobatto interpolation nodes for stability as discussed by Flad and Gassner in \cite{flad2017use}.
We investigate the influence of the discretization on the RL agent by computing LES at different resolutions as listed in \tabref{tab:les_configs}.
For this, we also employ two different polynomial degrees, i.e. $N=5$ and $N=7$, which results in two different discretizations. 
The maximum wavenumber used for optimization $k_{max}$ is then chosen for each case according to the respective resolution employed.
Theoretically, a minimum of $n_{ppw}=\pi$ points per wavelength is required to resolve a wavenumber accurately with a polynomial basis.
For the polynomial degrees employed here, the resolution capabilities can be estimated as $n_{ppw}\approx 4$ \cite{gassner2011comparison}.
However, to increase the difficulty of the optimization task, we choose $k_{max}$ such that $2.6 \le n_{ppw} \le 3$.
This forces the RL algorithm to optimize over all represented wavenumbers.
 
All simulations are initialized with flow states that are obtained by projecting the high-fidelity DNS solution, which is obtained a priori, onto the resolution of the respective environment using an $L_2$-projection.
The forced HIT case has a Reynolds number of $\mathrm{Re}_{\lambda}\approx 180$ with respect to the Taylor microscale.
The initial states used for training are drawn from the filtered DNS evaluated at $t_{DNS}=3,4,5,6$ and a single flow state at $t_{DNS}=8$ is kept hidden for testing, i.e. to evaluate the policy's performance on unseen data.
The LES are then simulated for $\Delta t_{end}=5$ using the FLEXI solver, while the elementwise $C_s$ parameters are updated every $\Delta t_{RL}=0.1$.
The large-eddy turnover time with approximately $t'\approx 0.7$ acts as a characteristic timescale, which results in 7 characteristic timescales simulated in each LES run.
This is considered sufficient to incorporate long-term effects into the training process.

For the RL algorithm, we use the Adam optimizer \cite{kingma2014adam} with a learning rate of $10^{-4}$ for the policy and the value network.
All configurations were trained for 5 epochs in each training iteration, except the 48 DOF configuration, which was only trained for a single epoch.
Instead of using mini-batching, the gradients were computed with respect to all sampled experience.
The weighting factor between the policy and the value estimation loss is set to 0.5 and the clipping parameter to $\ppoclipping=0.2$.
No additional regularization was used and the originally proposed entropy regularization coefficient for PPO is set to zero.
Moreover, we chose a discount factor of $\gamma=0.995$.
To obtain stochastic predictions from the deterministic policy, we sample actions from a normal distribution which is determined by using the ANN's predictions as mean and a fixed standard deviation of 0.02, which corresponds to about 10\% of the theoretical $C_s=0.17$ suggested for Smagorinsky's model in the literature.

\section{Results}
\label{sec:results}

The following section discusses the training of the agents and their inference on unseen starting states.
Moreover, the influence of invariant input features on the training success is assessed.
Lastly, the generalization abilities of the trained models to other resolutions and a higher Reynolds number are investigated.%
\footnote{The trained models and the required training files can be obtained from \url{https://github.com/flexi-framework/DRL_LES}.}

\subsection{Training}
\label{sec:results_training}

\begin{figure}[htb!]
  \centering
  \includegraphics[width=0.99\linewidth]{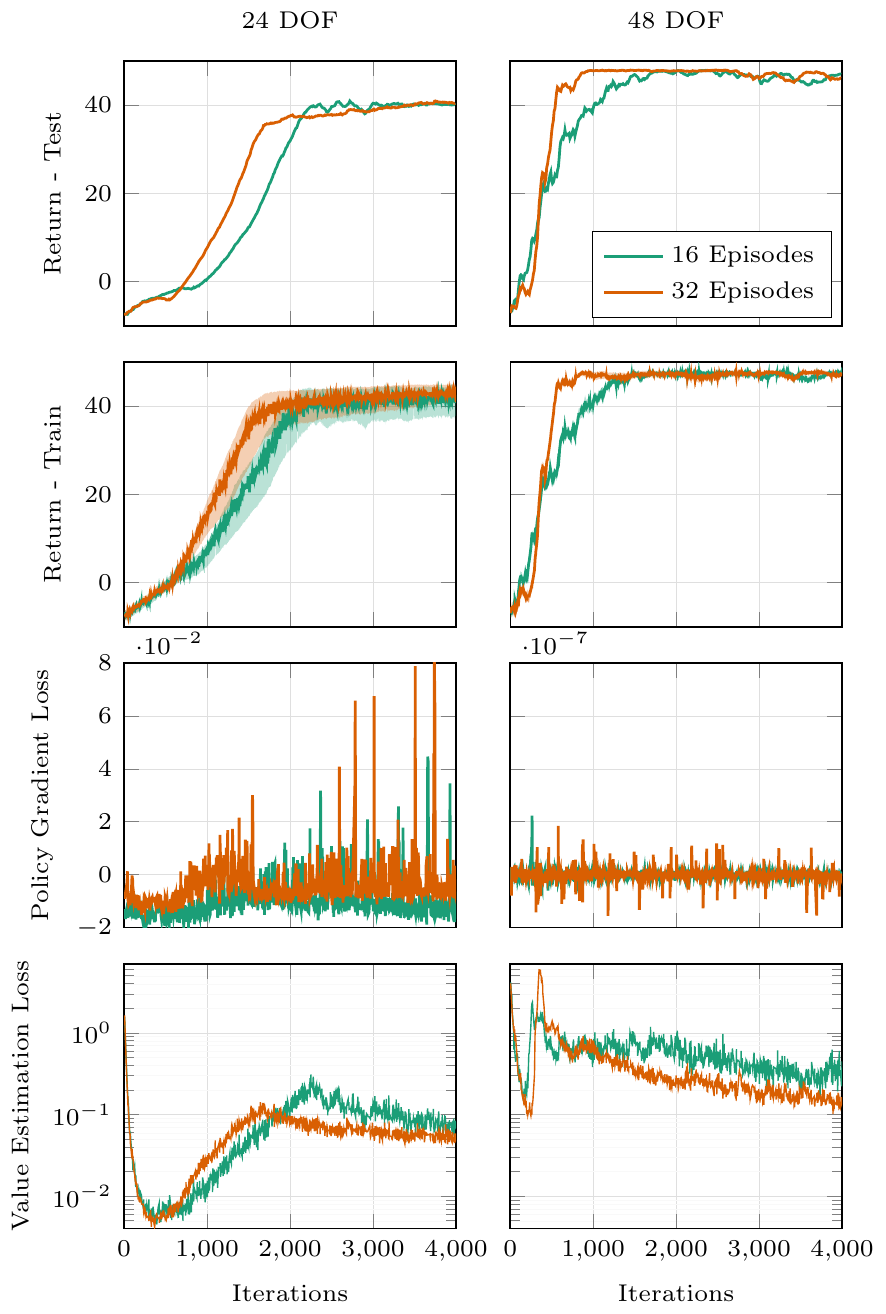}
  \caption{Training results for the 24 DOF (left) and 48 DOF (right) test case showing (from top to bottom) the undiscounted collected return on the unseen testing data, the same for the training episodes with the minimum and maximum return indicated by the shaded area, the policy gradient loss according to \eqref{eq:clipping_loss} and the value estimation loss for the value estimation ANN. The results are obtained for either 16 or 32 full episodes used per policy update. Please note that the y-axis of the policy gradient loss is scaled differently for both cases. Since the training simulations update the predictions each $\Delta t_{RL}=0.1$ and are simulated up to $t_{end}=5$, the maximum undiscounted return is $\return_{max}=50$ considering the reward in each step is normalized to $\reward_t\in[-1,1]$.}
  \label{fig:results_training}
\end{figure}

The training took between a single day for the smallest and up to 5 days for the largest cases on the HAWK supercomputer at the High-Performance Computing Center Stuttgart (HLRS).
The training was performed using up to 1024 CPU cores for the simulations and a single GPU node for model execution and training.
For more details on the hardware configuration, the reader is referred to \cite{kurz2022deep}.
The training behavior of the 24 DOF and the 48 DOF configurations is shown exemplarily in \figref{fig:results_training}, since these configurations constitute the lowest and highest employed resolution, respectively.
Both cases were trained once with 16 and with 32 episodes per parameter update.
For all cases, the collected return in the simulation is negative for the randomly initialized policy.
Starting from this initial random policy, the collected return increases during training until the return converges and plateaus just under the maximum undiscounted return of $\return_{max}=50$.
The maximum return follows from the 50 rewards collected during a simulation and the normalization of the reward function in \eqref{eq:reward} that guarantees $r_t\in[-1,1]$.
The convergence of the collected reward indicates that the RL algorithm has found a local optimum with its current policy.
Generally, the gradient estimator used for the gradient ascent algorithm should be more accurate if the amount of sampled episodes per parameter update is increased.
This can speed up the training, since a better approximated gradient can lead to more efficient parameter updates and thus reduce the overall training iterations needed for convergence.
This was indeed observed consistently for all investigated configurations.
Interestingly, the larger 48 DOF case requires less training iterations for convergence compared to the 24 DOF case.
Moreover, the 48 DOF case exhibits less variance in the return sampled in the training runs.
We attribute this reduced variance to two different factors.
First, the influence of the eddy-viscosity model on the overall flow decreases with increasing resolution, since the model accounts for a diminishing amount of unresolved kinetic energy in the flow.
Secondly, a single flow state contains more elements for the larger cases.
Since the policy trains on elementwise data, the same amount of episodes thus provides more training samples for the larger configurations.
The overall reduced variance in the training process then might cause more sample-efficient and faster optimization.
However, since the amount of required iterations for convergence did not decrease consistently with increasing resolution, the faster training might also be simply caused by the stochasticity of the training process.

The last row in \figref{fig:results_training} shows the loss of the value estimation ANN, which is trained to approximate the expected future return starting from a given state of the environment.
The value ANN is initialized with random weights and thus gives a poor estimate of the expected return in the beginning of the training, which results in a high initial loss.
The loss then decreases as the value ANN learns a first sensible estimate of the expected return.
Since the policy then starts to improve, i.e. to collect more reward, also the expected return and thus the training targets of the value ANN change.
Therefore, the value estimation loss increases as the policy improves, since the value ANN has to catch up constantly with the policy.
Once the return reaches a plateau, the expected return as target quantity for the training of the value ANN becomes more stable and the value estimation loss decreases again.

\subsection{Inference}
\label{sec:inference}

\begin{figure*}[htb!]
  \centering
  \includegraphics[width=\textwidth]{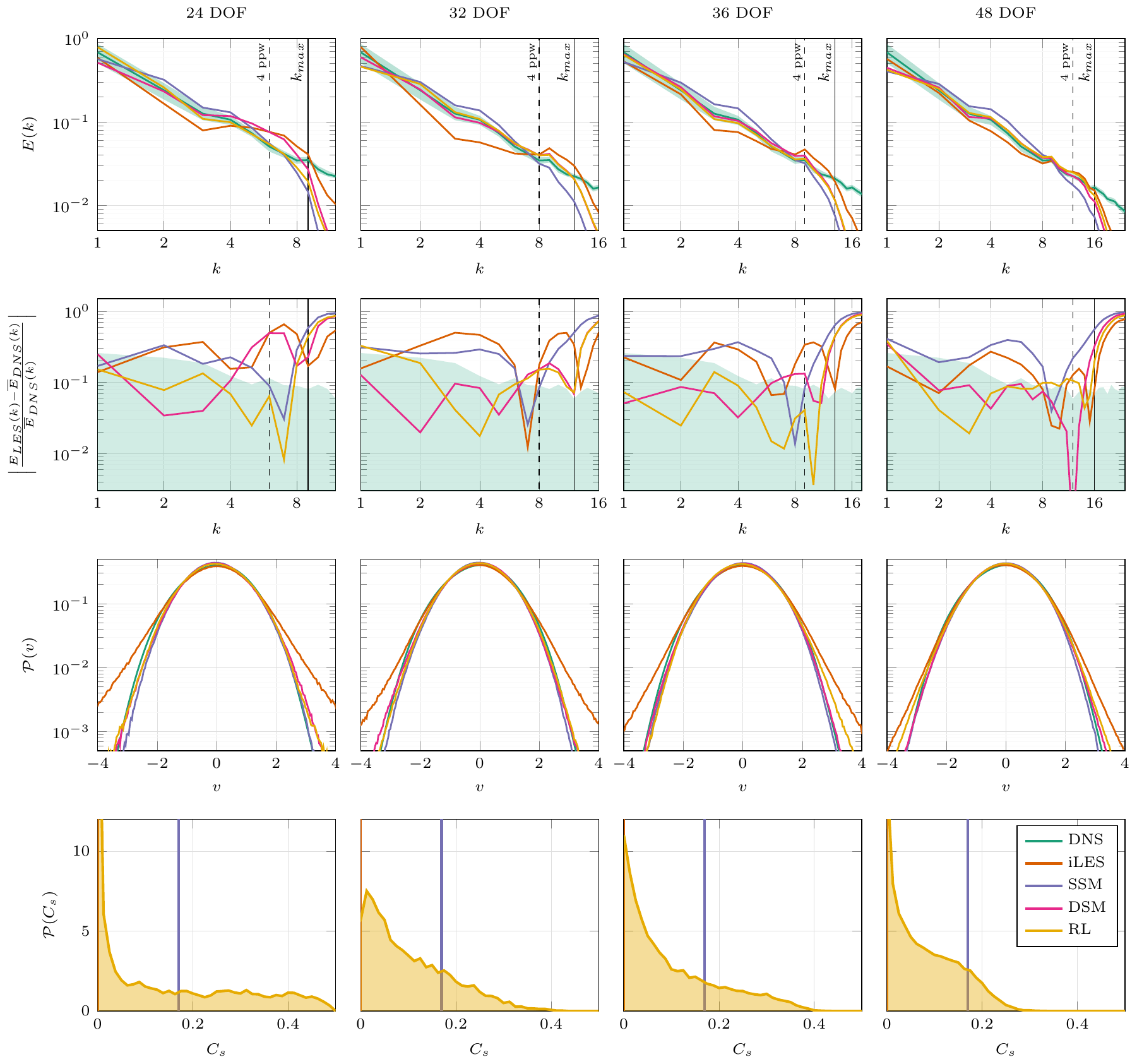}
  \caption{Results for the trained RL models (from left to right) in the 24 DOF, 32 DOF, 36 DOF and 48 DOF configuration averaged over $t\in[10,20]$ for an LES initialized with the unseen testing sample. Reported are (from top to bottom) the averaged energy spectra over the wavenumbers $k$, the relative error of the energy spectra with respect to the DNS solution, the distributions $\mathcal{P}(\cdot)$ of the velocity fluctuations, and the distribution of the predicted $C_s$ parameters. The results for the underlying DNS solution as well as an implicitly modeled LES (iLES), the SSM with $C_s=0.17$ and the DSM are shown for comparison. The shaded area for the DNS energy spectrum indicates the maximum and minimum amplitudes observed for each mode. The wavenumber that is resolved by the discretization with 4 points per wavelength is shown dashed and the maximum wavenumber used for optimization $k_{max}$ is indicated with a solid black line.}
  \label{fig:spectra_n5}
\end{figure*}

The performance of the trained RL models is evaluated based on an LES, which is initialized with the unseen testing state and is computed for $t_{end}=20$.
All results reported in the following are obtained and averaged over the timeframe of $t\in[10,20]$.
Since the models are trained only on simulations with $t_{end}=5$, this allows to assess the long-term behavior of the models and whether the simulation time used during training was sufficient.
The results in \figref{fig:spectra_n5} demonstrate that the RL models are indeed long-term stable.
To assess the accuracy of the RL models, they are compared to the SSM and DSM as well as the implicit model.
As could be expected, the implicitly modeled LES exhibits a buildup of energy in the upper wavenumbers due to lacking dissipation.
The SSM with $C_s=0.17$ on the other hand introduces too much dissipation and thus fails to preserve the wavenumbers near the resolution limit of the underlying numerical scheme.
The RL model clearly outperforms both models for all considered configurations by matching the target spectrum almost perfectly up to and even beyond the discretization's resolution limit of around 4 points per wavelength.
The advantage of the RL models is more pronounced for the small cases, where the turbulence model has more impact on the overall flow.
Interestingly, the errors for the different wavenumbers are distributed more evenly for the RL model.
This might stem from the objective of minimizing the squared error in the energy spectrum in order to increase the reward as given in \eqref{eq:reward}. 

The DSM, however, reproduces the DNS spectra with similar accuracy as the RL model.
This is to be expected, since the DSM is known to provide near perfect results for the HIT case, as long as the test filter is situated in the inertial range.
The RL model achieves a similar level of accuracy, i.e. a near optimal policy, without having access to the additional filtering procedure of the DSM.
It is interesting that the most notable difference between the DSM and RL is for the 24 DOF case, where the RL agent provides a signiﬁcantly better energy spectrum.
A likely explanation for this loss in accuracy of the DSM is that the LES resolution in the 24 DOF case is too coarse for the test filtering to occur in the scale-similar region.
Thus, no meaningful information is provided to the DSM procedure.
The RL model, however, seems to compensate for this lack of resolution and reproduces the DNS spectrum with very good accuracy.
It is important to stress here once again, that the underlying forcing of the test case prescribes the overall energy budget of the simulation.
Therefore, errors in the high wavenumbers might influence the energy contained in the low wavenumber and vice versa.
This stresses the capabilities of the RL models even more, which have learned to interact with this forcing such that the energy spectrum fits the prescribed one based on local information of the flow field only.

\begin{figure*}
  \centering
  \includegraphics[width=\textwidth]{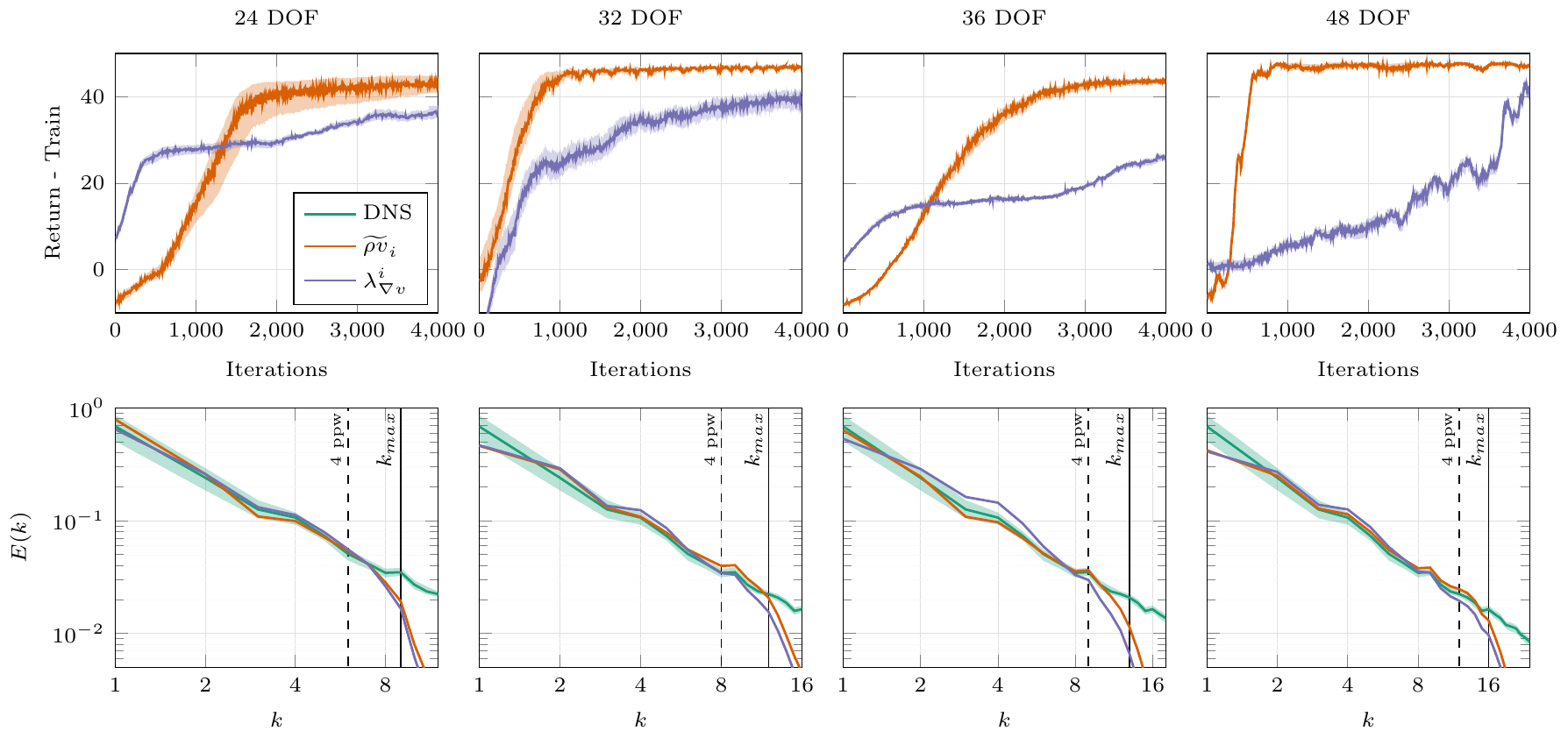}
  \caption{Comparison between the RL models trained with the local momentum field $\widetilde{\rho v}_i$ and with the invariants of the velocity gradient tensor $\lambda_{\nabla v}^i$ as inputs for the (from left to right) 24 DOF, 32 DOF, 36 DOF and 48 DOF configuration. The top row shows the undiscounted return collected during training with the shaded area indicating the episodes with the maximum and minimum return. The lower row shows the energy spectra in comparison to the DNS solution. The shaded area in the spectra indicates the maximum and minimum observed energy contained in the respective wavenumber during the DNS. The wavenumber that is resolved by the discretization with 4 points per wavelength is shown dashed and the maximum wavenumber used for optimization $k_{max}$ is indicated with a solid black line.}
  \label{fig:invariants}
\end{figure*}

It seems important to stress again that the energy spectra are the optimization target for the training and thus might provide only limited insight into the model's overall ability to reproduce the required turbulent statistics.
To this end, the distributions of the velocity fluctuations produced by the different models are investigated as additional important measure of the models' performance.
The differences between the models observed here are consistent with the obtained energy spectra.
Since additional dissipation tends to reduce velocity fluctuations, the SSM exhibits the narrowest and the implicit LES the broadest distribution.
For all cases, the velocity fluctuations produced by the RL model appear to be balanced between both effects, since they follow the DNS distribution more closely than the SSM for small fluctuations in the center of the distribution, but do not exhibit the overpronounced tails observed for the implicit LES.
However, the RL model produces an unsymmetrical distribution of velocity fluctuations for the 36DOF case.
This behavior was not observed for any other configuration or model and its origin is still subject of on-going investigations.
It is also unclear whether this behavior emerges since the symmetry condition has to be learned by the model implicitly during the training and might be not strict enough or whether this effect emerges from the long-term interactions between the agent and the forcing method.

The distributions of $C_s$ in the bottom row of \figref{fig:spectra_n5} show qualitatively similar results for all cases.
Most predictions are close to zero with an decreasing amount of higher values.
The overall range seems to strongly depend on the resolution, since the agent exploits almost all of its available action space for the 24 DOF case performing actions near the prescribed maximum of 0.5.
In contrast, the largest $C_s$ prediction for the 48 DOF case does not exceed 0.3.
The predictions' mean thus decreases for increasing resolutions as is consistent with the understanding of an increased LES resolution on the Smagorinsky model.
This difference might stem either from the policy itself, i.e. the policies learned indeed different distributions, or from the input data of the respective resolutions, which might exhibit different distributions.
Nonetheless, these results show the flexibility and capability of the RL training approach to incorporate physical constraints into the model through the choice of the input features.
We did not adapt the expressivity of the ANN between the two input selections, which might increase the performance.

Sarghini et al. \cite{sarghini2003neural} and Maulik et al. \cite{maulik2021deploying} reported a speedup by applying ANN instead of the computationally expensive DSM.
To this end, we compared the computational time required to evaluate the policy and the dynamic procedure of the DSM on a single CPU core for the 48 DOF case.
We found that the time required was comparable for both cases with the RL policy requiring around 10 per cent more time on our hardware.
These results are encouraging, since we did not perform any optimizations of the policy in terms of computational efficiency or model size and did not use GPU acceleration for this comparison, which improves the performance of the RL-based policy significantly.

\subsection{Input Features}
\label{sec:results_features}

In a next step, the models trained on the local momentum field as inputs are compared to the models trained on the five invariants of the velocity gradient tensor $\lambda_{\nabla u}^i$.
The results in \figref{fig:invariants} indicate that again all models successfully improve during training.
However, the training is generally slower and less stable than the former models.
As a result, the final models using the invariants as inputs still partly improve over the analytical models, but perform worse than the models using the momentum field, especially in the 36 DOF case.
This indicates that it is generally harder to learn a sensible policy from the invariants.
To investigate this further, we increased the training time.
While the models always seemed to improve to some degree, even with double the amount of training iterations the gap to former models stayed quite substantial.

We attribute this to the different distributions of the input quantities.
For the considered HIT test case, the velocity fluctuations have zero mean and a root-mean-squared (RMS) magnitude of unity by construction.
The fluctuations are thus approximately normally distributed with zero mean and unit variance.
This is the optimal distribution for input quantities in machine learning, which typically has to be achieved by normalizing the inputs accordingly.
Since the velocity fluctuations are intrinsically linked to the energy budget in the simulation and are thus constraint by the forcing, the agent's actions have only relatively limited impact on the distribution of velocity fluctuations, as already shown in \figref{fig:spectra_n5}.
In contrast, the invariants of the velocity gradient tensor typically span orders of magnitude.
Moreover, the computation of the gradients, and thus the distribution of the invariants, differ widely depending on the employed numerical discretization.
For DG, the gradients are intrinsically discontinuous across element faces and are also known to produce large gradients at the element faces for underresolved turbulence.
This is especially problematic for the initial states, which are obtained by projecting the DNS flow field onto the DG basis with respective LES resolution.
This projection causes large gradients and thus large values for the invariants, which makes it hard to normalize them to a tamer distribution.
We thus assume that the problems in the training stem from the gradient computation of the DG method, for which the gradients exhibit a complex distribution and the invariants computed from it span orders of magnitude, which makes training more difficult for the agent.

\subsection{Generalization to other Resolutions}
\label{sec:results_generalize}

\begin{figure}
  \centering
  \includegraphics[width=0.99\linewidth]{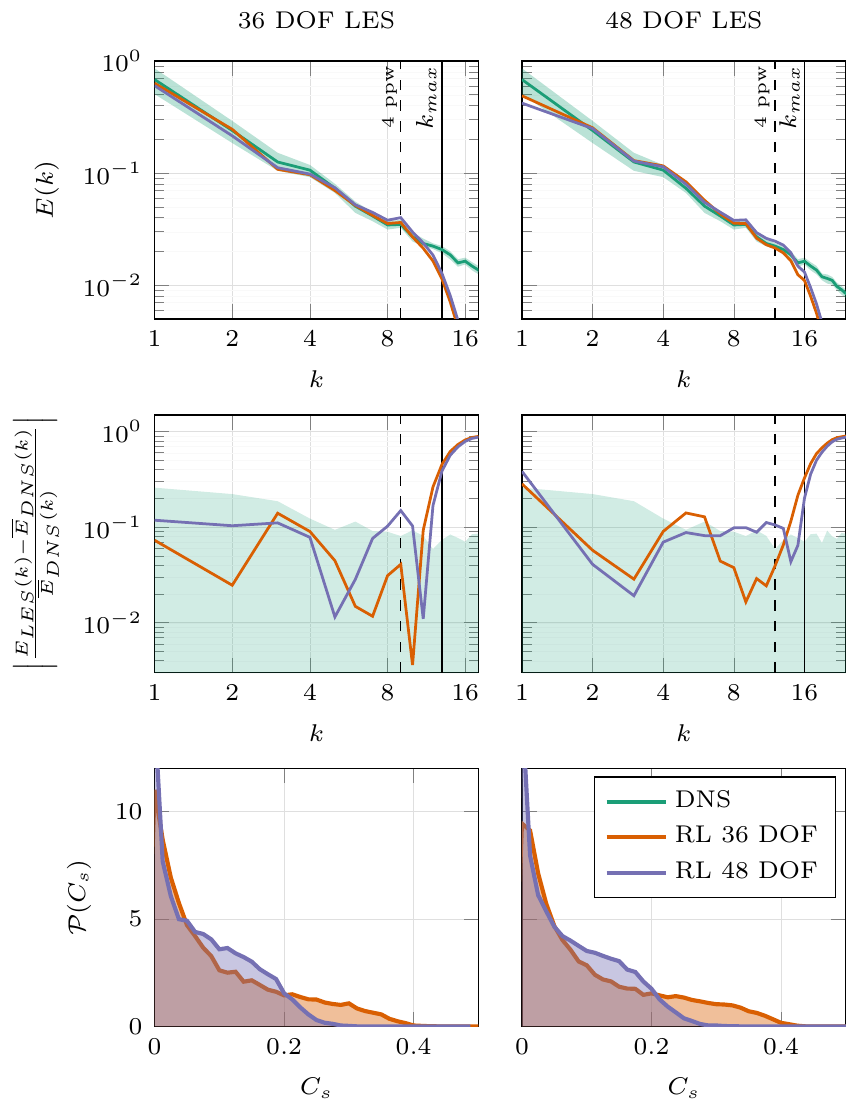}
  \caption{Results for the RL model trained on the 48 DOF evaluated on the 36 DOF case (left) and vice versa (right). Given are the energy spectra over the wavenumbers $k$ (top), the error of the spectra in comparison to the DNS solution (center) and the distribution of the predicted $C_s$ parameters (bottom). The shaded area indicates the maximum and minimum observed energy contained in the respective wavenumber during the DNS. The wavenumber that is resolved by the discretization with 4 points per wavelength is shown dashed and the maximum wavenumber used for optimization $k_{max}$ is indicated with a solid black line.}
  \label{fig:generalization_resolution}
\end{figure}

To demonstrate that the trained models can generalize to different resolutions, the model trained on the 48 DOF resolution is evaluated on the 36 DOF configuration and vice versa.
This allows to assess how well the trained models can be transferred to LES cases with either more or less resolution.
The results shown in \figref{fig:generalization_resolution} demonstrate that the trained models can also provide stable and accurate results in LES with different resolutions.
This is especially remarkable, since the policy's field of vision for the policy shrinks with increasing resolution due to the elementwise input and output quantities.
The RL model trained natively on the 48 DOF case shows a slight increase in energy in the higher wavenumbers, while the 36 DOF model seems to be slightly more dissipative.
However, the overall errors in the energy spectrum appear to be comparable for both cases.
The classical turbulence models are not shown for clarity.
However, since both RL models provide almost identical energy spectra, they still outperform the SSM and implicit LES model for both resolutions, while matching the performance of the DSM.

Also, the distribution of the models' predictions are almost identical for both cases and thus do not appear to change depending on the LES resolution.
The predictions of the 36 DOF model exhibit a much wider tail, with a maximum prediction of around $C_{s,max}=0.4$.
In contrast, the predictions of the trained 48 DOF model do not exceed $C_{s,max}=0.3$.
Interestingly, the models are still able to reproduce the target energy spectrum despite the deviations in their policies.
It is plausible to assume that the models will generalize even better, if they are trained on a variety of different resolutions, instead of only a single one.
These pronounced differences in the learned policies indicate that the distribution of predictions is not only induced by the input features but is a characteristic property of the policy trained on the respective resolution and the employed discretization.
This again demonstrates that the different discretizations induce different implicit LES filters, which again require different policies to match the underlying energy spectrum.
Thus, the proposed framework allows to develop discretization-adapted turbulent models for implicit LES.

\subsection{Generalization to other Reynolds Numbers}
\label{sec:results_generalize_re}

In a final step, we demonstrate that the trained RL policy is able to generalize to higher Reynolds numbers.
For this, the trained agents for the different resolutions are applied to a HIT flow at a Reynolds number of $Re_{\lambda}\approx 240$, which is considerably higher than the Reynolds number $Re_{\lambda}\approx 180$ used for training.
Analogously to \secref{sec:inference}, the LES were initialized from filtering the DNS flow field at a random point in time to the required resolution.
The LES was then advanced in time for $t_{end}=20$ and the results were averaged over the timeframe of $t\in [10,20]$ to investigate the long term effects of the model onto the flow.

\begin{figure*}[htb!]
  \centering
  \includegraphics[width=\textwidth]{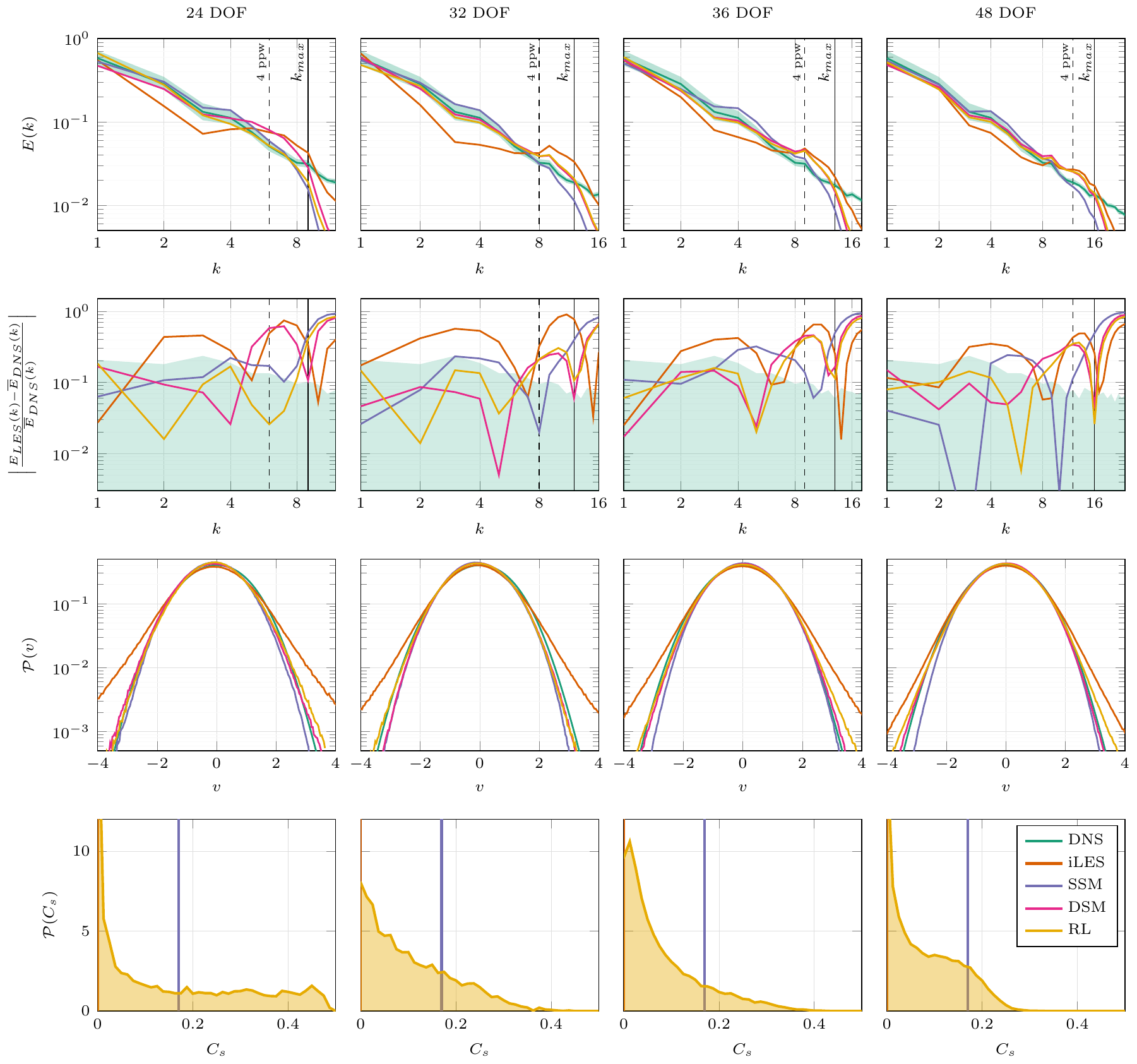}
  \caption{Results for the RL models trained on $Re_{\lambda}\approx180$ evaluated on a HIT flow with $Re_{\lambda}\approx240$. Reported are (from top to bottom) the averaged energy spectra over the wavenumbers $k$, the relative error of the energy spectra with respect to the DNS solution, the distributions $\mathcal{P}(\cdot)$ of the velocity fluctuations, and the distribution of the predicted $C_s$ parameters. The results for the underlying DNS solution as well as an implicitly modeled LES (iLES), the SSM with $C_s=0.17$ and the DSM are shown for comparison. The shaded area for the DNS energy spectrum indicates the maximum and minimum amplitudes observed for each mode. The wavenumber that is resolved by the discretization with 4 points per wavelength is shown dashed and the maximum wavenumber used for optimization $k_{max}$ is indicated with a solid black line.}
  \label{fig:generalization_re}
\end{figure*}

The results in \figref{fig:generalization_re} indicate that the trained models can indeed generalize to flows at higher Reynolds numbers.
Most importantly, the RL models still provide long-term stable simulations.
The RL models show similar behavior as for the Reynolds number seen during training. 
For the 32 DOF, 36 DOF and 48 DOF cases the RL models is able to reproduce the energy spectrum more accurately than the implicit model and the SSM, but with similar accuracy as the DSM. 
Interestingly, the DSM and RL models exhibit a similar buildup of energy near the cutoff wavenumber, which might indicate that these models lack sufficient dissipation.
Moreover, the RL model sill outperforms the other models and especially the DSM for the 24 DOF simulation, where the modeling assumptions of the SSM and DSM most probably do not hold.

These results are very promising, since they indicate that the trained RL policies are able to extrapolate to other Reynolds numbers (at least to a moderate extent).
The trained policies are thus able to generalize to higher Reynolds number flows as well as other LES resolutions, while matching or even improving on the performance of the DSM, which is known to provide outstanding results for HIT flows.

\section{Conclusions}
\label{sec:conclusion}

In this work, we have successfully applied the RL framework Relexi \cite{kurz2022deep} for the development of data-driven turbulence models for implicitly filtered LES.
To this end, we have formulated the turbulence modeling task as an RL problem using LES of forced HIT as the training environment.
The LES was filtered implicitly by a modern high-order discretization based on a DG scheme to demonstrate that the RL framework can incorporate complex implicit LES filter functions into the training process by design.
The employed policy is based on a CNN architecture which relies on elementwise inputs and outputs only without the need of global flow statistics.
The agent was trained within the Relexi framework using the PPO algorithm to adapt the elementwise model parameters of Smagorinsky's model dynamically in space and time.
The underlying spectrum of turbulent kinetic energy of a high-fidelity solution was used as optimization target, which the agent should reproduce as accurate as possible.
We have demonstrated that the agent is able to learn highly accurate and long-term stable turbulence models, which clearly outperform established analytical models.
Moreover, the RL-based models are able to reproduce the target spectrum in actual simulation up to and beyond the expected resolution capabilities of the underlying discretization.
The models were shown to generalize well to other resolutions and other Reynolds numbers they were not originally trained on.
In summary, we have demonstrated that RL provides a framework for consistent, accurate and stable turbulence modeling for implicitly filtered LES.

Future work will focus on the incorporation if physical invariances into the models.
While employing the invariants of the gradient tensor as input features for the policy network incorporates Galilean invariance in a pointwise fashion, the CNN architecture is by default only invariant to shifting operations but not to rotations or reflections.
Since those invariances are not incorporated into the architecture, they have to be learned implicitly from data, which reduces the sample efficiency of the training.
Further investigations will incorporate the physical knowledge about the invariances of the underlying equations into the model architecture with specialized architectures like group convolutional layers \cite{cohen2016group}, which were already applied successfully for turbulence modeling in \cite{pawar2022frame} and in \cite{guan2023learning}, or by interpreting the input data as a graph containing only neighboring information without explicit orientation \cite{niepert2016learning,wu2020comprehensive}.
In addition, the framework will be applied to more sophisticated test cases like wall-bounded flows or shear flows.
In this context, also the use of more complex actions spaces for the agent will be investigated.

To summarize, our work here serves as a proof of concept for finding optimal LES closure models through reinforcement learning.
We do not propose to use any of the models prescribed here in practice, but our results demonstrate how to develop near optimal models with this approach, where the physical consistency and expert knowledge in model creation is fused with the currently most powerful learning algorithm for dynamical systems.
Beyond the application to LES modeling, our work here demonstrates the potential of a high-fidelity flow solver in the loop with an RL optimizer.

\section*{Acknowledgment} %
The research presented in this paper was funded by Deutsche Forschungsgemeinschaft (DFG, German Research Foundation) under Germany's Excellence Strategy - EXC 2075 - 390740016.
The authors gratefully acknowledge the support and the computing time on "Hawk" provided by the HLRS through the project "hpcdg" and the support by the Stuttgart Center for Simulation Science (SimTech).
\appendix

\section{Dynamic Smagorinsky Model}
\label{app:dynsmago}

Germano et al. proposed in \cite{germano1991dynamic} an improvement to the standard Smagorinsky model by setting the Smagorinsky constant $C_s$ not to a fixed value set a priori, but rather to adapt it dynamically in space and time dependent on the current state of the flow.
This dynamic procedure is based on applying a test filter $\widehat{(\cdot)}$ to the solution.
The LES solution can then be used to compute the resolved stress tensor
\begin{equation}
  L_{ij} = \widehat{\widetilde{u}_i \widetilde{u}_j} - \widehat{\widetilde{u}}_i \widehat{\widetilde{u}}_j,
\end{equation}
which entails the subgrid stresses induced by the test filter.
Germano's identity finds a correlation between those resolved subgrid stresses and the unknown subgrid stress induced by the LES filter.
Lilly \cite{lilly1992proposed} proposed a least-squares approach to compute $C_s^2$ such that Germano's identity is optimally obeyed, which gives
\begin{equation}
  C_s^2 = \frac{\langle L_{ij}M_{ij}\rangle_{avg}}{\langle M_{ij}M_{ij} \rangle_{avg}}.
  \label{eq:dynsmago_ls}
\end{equation}
Here, $\langle\cdot\rangle_{avg}$ denotes some averaging operation to avoid extremely large coefficients and thus to stabilize the model.
The tensor $M_{ij}$ is shorthand for the expression
\begin{equation}
  M_{ij} = 2 \Delta^2 \widehat{\sqrt{2\tilde{S}_{kl}\tilde{S}_{kl}}\,\tilde{S}_{ij}}  - 2 \widehat{\Delta}^2 \sqrt{2\widehat{\tilde{S}}_{kl}\widehat{\tilde{S}}_{kl}}\, \widehat{\tilde{S}}_{ij},
\end{equation}
with $\widehat{\Delta}$ as the filter width of the test filter and with the rate-of-strain tensor on the test filter level $\widehat{\tilde{S}}$ computed analogously to \eqref{eq:smago} based on the test-filtered velocity field $\widehat{\tilde{u_i}}$.

In the DG context, the dynamic procedure is applied in an elementwise fashion.
As test filter $\widehat{(\cdot)}$ with filter width $\widehat{\Delta}$, we apply a modal cut-off filter to the solution polynomial with a degree of $N_{test}=2$ and $N_{test}=3$ for the cases employing a polynomial degree of $N=5$ and $N=7$, respectively.
Moreover, the averaging operator $\langle\cdot\rangle_{avg}$ for the numerator and denominator in \eqref{eq:dynsmago_ls} is chosen as an average in each element yielding a single coefficient $C_s^2$ for each element.
The resulting eddy-viscosity is then clipped to the range $\mu_t \in [-100\mu,1000\mu]$ with respect to the physical viscosity, since the unclipped model produced from time to time a few huge values for $C_s^2$, which deteriorated the solution quite drastically.
The clipping range itself seemed to have only very limited influence on the model's behavior in our tests and provided very similar results for a wide range of different clipping intervals.

\bibliographystyle{elsarticle-num}
\bibliography{bibliography}

\end{document}